\documentclass[prd,aps,twocolumn,floatfix,preprintnumbers,groupedaddress,nofootinbib]{revtex4-1}
\pdfoutput=1

\usepackage{amsmath}
\usepackage{amssymb}
\usepackage{graphicx}
\usepackage{float}
\usepackage{braket}
\usepackage{hyperref}
\usepackage{pbox}
\usepackage{tabularx}
\usepackage{enumitem}
\usepackage{tikz}\usetikzlibrary{intersections}
\usepackage{epstopdf}
\def\beq{\begin{equation}}
\def\eeq{\end{equation}}
\def\beeq{\begin{eqnarray}}
\def\eeeq{\end{eqnarray}}

\def\gp2{g^{\prime 2}}

\def\lra#1{\overset{\text{\scriptsize$\leftrightarrow$}}{#1}}

\newcommand{\pt}{\partial}

\newcommand{\lag}{\mathcal{L}}
\newcommand{\Tr}{\mathrm{Tr}}
\newcommand{\abs}[1]{\lvert #1 \rvert}

\newcommand{\Gev}{\, \mathrm{GeV}}
\newcommand{\Tev}{\, \mathrm{TeV}}

\usepackage{color}
\begin{document}   
\preprint{DAMTP-2015-32}
\preprint{Cavendish-HEP-15/05}
\title{Anatomy of the ATLAS diboson anomaly} 
\date{\today}
\author{B.~C.~Allanach} 
\email{B.C.Allanach@damtp.cam.ac.uk}
\affiliation{Department of Applied Mathematics and Theoretical Physics, Centre
  for Mathematical Sciences, University of Cambridge, Wilberforce Road,
  Cambridge, UK} 
\author{Ben Gripaios} 
\email{gripaios@hep.phy.cam.ac.uk}
\author{Dave Sutherland}
\email{dws28@cam.ac.uk}
\affiliation{Cavendish Laboratory, University of Cambridge, J.J. Thomson Ave,
  Cambridge, UK} 

\begin{abstract} 
We perform a general analysis of new physics interpretations of the
recent ATLAS diboson excesses over Standard Model expectations in LHC Run I
collisions. 
Firstly, we estimate a likelihood
function in terms of the truth signal in the $WW$, $WZ$, and $ZZ$
channels, 
finding that the maximum has zero events in the $WZ$ channel,
though the likelihood is sufficiently flat to allow other scenarios. Secondly,
we survey the possible 
effective field theories containing the Standard Model plus a new
resonance that could explain the data, identifying two possibilities,
{\em viz.}\/ a vector that is either a left- or right-handed $SU(2)$
triplet. Finally, we compare these models with other experimental 
data and determine the parameter regions in which they provide a consistent
explanation. 
\end{abstract}   
\maketitle 

\section{Introduction \label{sec:intro}}
The ATLAS experiment has recently reported~\cite{Aad:2015owa} three excesses
in 
searches based on jet substructure methods for resonances decaying into 
dibosons (where each jet is interpreted as being a hadronically decaying
$Z^0$ boson or $W^\pm$ boson). 

The excesses appear at a diboson invariant mass of around 2
TeV
in each of
three decay channels studied -- $WZ$, $WW$, and
$ZZ$ -- and have local significances of 3.4, 2.6, and
2.9$\sigma$, respectively and a global significance of 2.5$\sigma$ in the
$WZ$ channel for an integrated luminosity ${\mathcal L}=20.3$ fb$^{-1}$. 
ATLAS
provided limits upon 
models that could produce such signals and showed that a 2 TeV
$W'$ with weak-boson size gauge coupling or a 2 TeV type I Randall-Sundrum (RS)
graviton both have production cross-sections too small to explain the apparent
excess~\cite{Aad:2015owa}. At the same time, CMS finds a global excess of
1.9$\sigma$ in a boosted search for $WH$, with the Higgs $H$ decaying
hadronically~\cite{CMS-PAS-EXO-14-010} and the $W$ decaying leptonically.

In this note, we explore possible new physics
interpretations of the ATLAS excesses. We note that there are other smaller
(below 
2$\sigma$) excesses 
in other searches for diboson
resonances~\cite{CMS-PAS-EXO-14-010,Khachatryan:2014gha,Aad:2015ufa,Khachatryan:2014hpa},
but we 
concentrate here 
on the ATLAS ones because they are the most statistically significant. We will
however, apply constraints from other searches in order to ensure that our new
physics explanations are not already excluded.
Interpreting the ATLAS excesses in terms of 
a resonance, data indicate that it is fairly narrow, with a width of less than
100 GeV or so. 
There have been some recent suggestions for such resonances: for instance, in
Ref.~\cite{Fukano:2015hga}, walking technicolor was invoked in order to
interpret the apparent 2 TeV resonance as a technirho (the discovered Higgs
boson is interpreted as a technidilaton). 
Several other works concentrate on $W'$ or $Z'$ vector bosons~\cite{Alves:2015mua,Hisano:2015gna,Cheung:2015nha,Xue:2015wha,Dobrescu:2015qna,Aguilar-Saavedra:2015rna,Gao:2015irw,Cao:2015lia,Cacciapaglia:2015eea}. Refs.~\cite{Franzosi:2015zra,Thamm:2015csa} also
have vector resonances motivated by composite dynamics. 
Ref.~\cite{Brehmer:2015cia} postulates a left-right symmetric model to
generate the necessary extra bosons. 

Here, we pursue a different approach. Rather than examining specific
  models, we
survey the possible models (by which we mean effective field theories,
valid at TeV scales) containing the Standard Model (SM) plus
a new resonance that can describe the ATLAS anomalies without gross conflict
with 
other data. To do so, we first
calculate a likelihood
for the truth distribution of events in the $WW$, $WZ$, and
$ZZ$ channels.\footnote{A likelihood analysis was carried out in
Ref.~\cite{Brehmer:2015cia}, but ours differs in several ways.}
Secondly, we use the likelihood analysis and other data to
pin down the qualitative features of a possible new physics model.
We argue that models based on an $SU(2)_L$ or $SU(2)_R$ triplet
vector are most plausible, both of which have already been exploited in
specific cases in the literature. However, our approach yields results that
are less model dependent.
\section{Likelihood analysis \label{sec:like}}
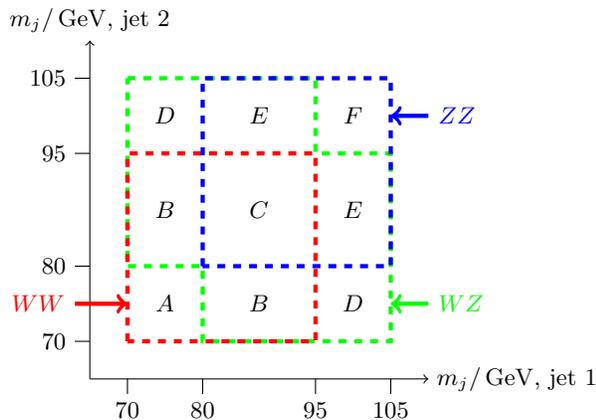
\begin{figure}
\centering
\begin{tikzpicture}[x=0.1cm,y=0.1cm]
    \draw[->] (65,65) -- coordinate (x axis mid) (110,65) node [right] {$m_j/\Gev$, jet 1};
    \draw[->] (65,65) -- coordinate (y axis mid) (65,110) node [above] {$m_j/\Gev$, jet 2};
    \foreach \x in {70,80,95,105}
        \draw (\x,65) -- (\x,63)
            node[anchor=north] {$\x$};
    \foreach \y in {70,80,95,105}
        \draw (65,\y) -- (63,\y)
            node[anchor=east] {$\y$};
    \draw[dashed,ultra thick,green] (70,105) --(95,105) -- (95,95) -- (105,95) -- (105,70) -- (80,70) -- (80,80) -- (70,80) -- (70,105);
    \draw[dashed,ultra thick,red] (70,70) --(70,95) -- (95,95) -- (95,70) -- (70,70);
    \draw[dashed,ultra thick,blue] (80,80) --(80,105) -- (105,105) -- (105,80) -- (80,80);

    \draw[->,ultra thick,red] (63,75) node [left,red] {$WW$} -- (70,75);
    \draw[->,ultra thick,blue] (110,100) node [right,blue] {$ZZ$} -- (105,100);
    \draw[->,ultra thick,green] (110,75) node [right,green] {$WZ$} -- (105,75);

    \node at (75,75) {$A$};
    \node at (87.5,75) {$B$};
    \node at (100,75) {$D$};
    \node at (75,87.5) {$B$};
    \node at (87.5,87.5) {$C$};
    \node at (75,100) {$D$};
    \node at (87.5,100) {$E$};
    \node at (100,100) {$F$};
    \node at (100,87.5) {$E$};

\end{tikzpicture}
\caption{A simple picture of the $WW$ (red), $WZ$ (green), and $ZZ$ (blue) signal regions used in
  \cite{Aad:2015owa}, in the $m_j$--$m_j$ plane of the two fat jets in an
  event. We also show our
  labelling of  
  disjoint signal regions $A,B,C,D,E,F$.\label{fig:sigreg}}
\end{figure}
The ATLAS diboson search \cite{Aad:2015owa} looks for evidence of a heavy
resonance which decays to $WW$, $WZ$, or $ZZ$, all of which subsequently decay
hadronically. From events containing two fat jets, jet substructure techniques
are used to select events wherein each fat jet contains two `prongs'
characteristic of a boosted $W$ or $Z$ decay, and thereby also to provide an
estimator $m_j$ of the invariant mass of the two subjets within each fat
jet. A fat jet is a `$W$ jet' if $69.4 \Gev < m_j  < 95.4 \Gev$; it is a `$Z$
jet' if $79.8 \Gev < m_j  < 105.8 \Gev$. The three signal regions are then
defined as a) $WW$, containing two $W$ jets; b) $WZ$, containing a $W$ jet and
a $Z$ jet, and c) $ZZ$, containing two $Z$ jets. Additionally, data are reported in the auxiliary information of \cite{Aad:2015owa} (\url{https://twiki.cern.ch/twiki/bin/view/AtlasPublic/ExoticsPublicResults}) for the combinations $WW+ZZ$ and $WW+WZ+ZZ$ of the three aforementioned regions. Clearly, a selected event may
be common to more than one signal region --- Figure~\ref{fig:sigreg} shows a
cartoon of the overlap of the signal regions in the $(m_j,m_j)$ plane of the
two fat jets.

It is not reported which of the excess events around $2 \Tev$ are common to more than one signal region; however, from the data available, we may infer the number of common excess events as follows. Hereafter, we shall concern ourselves with only the three bins of $m_{jj}$ nearest to 2 TeV, where ATLAS observed the
excesses.

Firstly, we seek to disentangle the overlapping signal regions into regions
that partition the parameter space of interest. We define six disjoint regions
$A$ to $F$ in the $(m_j,m_j)$ plane (Fig.~\ref{fig:sigreg}) which in
combination comprise the five ATLAS signal regions on which we have data:
\begin{align*}
 WW = A + B + C, \\
 ZZ = C + E + F, \\
 WZ = B + C + D + E, \\
 WW+ZZ = A + B + C + E + F, \\
 WW+WZ+ZZ = A + B + C + D + E + F.
\end{align*}
In Table~\ref{tab:eventNumbers} we show the three possible arrangements of the
events in the disjoint regions $A$--$F$ that are compatible with the ATLAS
data in the five overlapping regions, summed over the three $m_{jj}$ bins of
interest. In each of the five signal regions, ATLAS also provides an estimate
of the SM background by fitting a smooth curve to the observed $m_{jj}$
spectrum. There is a continuum of possible values for the SM background in the
six disjoint regions that are consistent with ATLAS's numbers in the five
overlapping regions --- we break the degeneracy by taking the solution with
equal ratios of background in $A$ to $F$, and in $B$ to $E$, as is consistent
with a QCD dijet background that is roughly flat in the $(m_j,m_j)$ plane. The
sums over the three $m_{jj}$ bins of the resulting expected values in the
regions $A$--$F$ are also shown in Table~\ref{tab:eventNumbers}. Note that the
uncertainties in the fitted background spectra are somewhat difficult to take
into 
account since they are likely to be correlated between the different channels,
and we do not have access to the correlation matrix. Fortunately, the
uncertainties are small and we neglect them.

\begin{table}
\begin{tabular}{ccccccc}
& $A$ & $B$ & $C$ & $D$ & $E$ & $F$ \\ \hline
$n^{\text{obs},1}_i$ & 2 & 6 & 5 & 0 & 4 & 0 \\
$n^{\text{obs},2}_i$ & 1 & 7 & 5 & 0 & 3 & 1 \\
$n^{\text{obs},3}_i$ & 0 & 8 & 5 & 0 & 2 & 2 \\
$\mu^\text{SM}_i$ & 2.09 & 2.72 & 1.00 & 2.43 & 0.46 & 0.34
\end{tabular}
\caption{The three possible arrangements of the observed events into the six
  disjoint signal regions $A$--$F$ of Fig.~\ref{fig:sigreg}, as well as our estimate
  of the expected event numbers in each region, summed  
  over the 
  three bins $m_{jj}/\textrm{TeV} \in [1.85-1.95,\
1.95-2.05,\ 2.05-2.15]$. \label{tab:eventNumbers}}
\end{table}

We now construct our likelihood fit to the LHC production cross section
times branching ratios ($\sigma \times \text{BR}$s) of a putative resonance
that is responsible for the excess events in
Table~\ref{tab:eventNumbers}. From \cite[Fig. 1c]{Aad:2015owa}, the
probabilities that the $W$ or $Z$ from a 
$2 \Tev$ diboson resonance has an $m_j$ in the $W$ or $Z$ window are
approximately as in Table~\ref{tab:wz}. Note that these numbers come
from the ATLAS simulation of a Randall-Sundrum graviton, which (when it decays
to $W$s or $Z$s) decays almost
exclusively to 
longitudinally polarised bosons; transversely polarised bosons would
have different $m_j$ distributions \cite{CMS-PAS-JME-13-006}, so the 
numbers should be taken {\em cum grano salis}\/ for other
new physics models.
\begin{table}\begin{center}
\begin{tabular}{c | c c c}
 & $W$ jet tag only & $W$ and $Z$ jet tag & $Z$ jet tag only\\ \hline
true $W$ & 0.25 & 0.36 & 0.04 \\
true $Z$ & 0.11 & 0.39 & 0.21
\end{tabular}
\caption{\label{tab:wz} Probability that a $W$ or $Z$ is tagged with a $W$ or
  $Z$ tag.}
\end{center}
\end{table}
The probabilities of a diboson resonance event satisfying the respective $m_j$
cuts of the signal regions $A$ to $F$ are thus shown in Table~\ref{tab:dib}, forming a 3 by 6 matrix $M_{ji}$.
\begin{table}\begin{center}
\begin{tabular}{c | c c c c c c}
$M_{ji}$ & $A$ & $B$ & $C$ & $D$ & $E$ & $F$ \\ \hline
true $WW$ & 0.063 & 0.182 & 0.132 & 0.018 & 0.025 & 0.001 \\
true $WZ$ & 0.028 & 0.139 & 0.143 & 0.057 & 0.090 & 0.007 \\
true $ZZ$ & 0.012 & 0.087 & 0.155 & 0.047 & 0.165 & 0.044 \\
\end{tabular}
\caption{Probability of different diboson candidates from a 2 TeV resonance
  being tagged in each signal region.\label{tab:dib}}\end{center}
\end{table}
We multiply the probabilities in Table~\ref{tab:dib} by a factor of
$\epsilon=0.33\times 0.67$ to 
match the 
reported efficiencies of \cite[Fig. 2b]{Aad:2015owa}, with an additional
probability of $0.67$ for the signal to be in the three $m_{jj}$ bins that we consider \cite[Fig. 2a]{Aad:2015owa}.  
Given a vector $s_j=\{s_{WW},\ s_{WZ},\ s_{ZZ}\}$ of the number of ``truth''
signal diboson pairs issuing from a putative 2 TeV resonance, 
we expect 
\begin{equation}
\mu_i = \mu_i^{SM}+ \sum_{j=1}^3 \epsilon b_j s_j M_{ji} \label{eq:mui}
\end{equation}
events to be tagged in
each signal region $i \in \{A,B,C,D,E,F\}$. $b_j=\{ 0.45,\ 0.47,\ 0.49\}$ are the totally hadronic
branching fractions of the diboson pairs.


We construct the joint likelihood of tagging $n_i$ events in each of the six signal regions:
\begin{eqnarray}
p(\{n_i\} | \{\mu_i \}) = \prod_{i \in \{A,B,C,D,E,F\} } P(n_i | \mu_i ).  \label{eq:probs}
\end{eqnarray}
The six probabilities on the right hand side of Eq.~\ref{eq:probs} are
Poissonian, i.e.\
\begin{equation}
P(n | \mu) = \frac{e^{-\mu} \mu^n}{n!}. \label{eq:poiss}
\end{equation}
Substituting Eqs.~\ref{eq:mui},\ref{eq:poiss} into Eq.~\ref{eq:probs}, we
obtain our likelihood 
\begin{eqnarray}
&&p(\{n_i^{\textrm{obs},\alpha}\} | 
s_{WW},\ s_{WZ},\ s_{ZZ}) = \nonumber \\
&& \sum_{\alpha=1}^3 \frac{\exp \left[ -\sum_{i \in \{A,B,C,D,E,F\} } \left(\mu_i^{SM} + \epsilon \sum_{j=1}^3 b_i s_j M_{ji}
\right)\right]}{\prod_{i \in \{A,B,C,D,E,F\} } n_{i}^{\textrm{obs},\alpha}! } \times \nonumber \\
&&\prod_{i \in \{A,B,C,D,E,F\}} \left( \mu^{SM}_{i} + \epsilon \sum_{j=1}^3 b_i
s_j M_{ji}\right)^{n_{i}^{\textrm{obs},\alpha}}, \label{eq:lk}
\end{eqnarray}
which includes the correlations coming from overlaps in the $W$ and $Z$
tags. We sum over the three independent partitions of events into the
regions $A$ to $F$ that are compatible with the ATLAS data, as labelled by
$\alpha$. Eq.~\ref{eq:lk} allows us to further investigate what the ATLAS fat
jet 
analysis dictates about the different decay channels for a signal. 
We turn the likelihood 
into a more familiar value of $\chi^2$ by 
$\chi^2=-2 \log p(\{n_i^{\textrm{obs},\alpha}\}| 
s_{WW},\ s_{WZ},\ s_{ZZ})$. Best-fit points will be found by minimising
$\chi^2$ (or, equivalently, maximising the likelihood).
We shall phrase our results in terms of 
the 
production cross section of the 2 TeV resonance $X$ times branching ratio for
each decay channel: $\sigma(X) \times BR(X \rightarrow i)=s_j / \mathcal{L}$.

Minimising $\chi^2$ over $s_j$, we obtain our best-fit point $s_j=\{ 106,\ 0,\ 118\}$, i.e.\
$\sigma(X^0) \times BR(X^0 \rightarrow {W^+W^-})=5.2$ fb,
 $\sigma(X^\pm) \times BR(X^\pm \rightarrow {W^\pm Z})=0$ fb, 
$\sigma(X^0) \times BR(X^0 \rightarrow {ZZ})=5.8$ fb corresponding to expected event numbers $\mu_{WW}=13.0$, $\mu_{WZ}=16.1$ and $\mu_{ZZ}=8.1$ in the three respective ATLAS signal regions $WW$, $WZ$, and $ZZ$. However, as we
shall show, the 
statistical uncertainties are such that sizeable deviations from this best-fit
point are possible.

\begin{figure}
\includegraphics[width=\columnwidth]{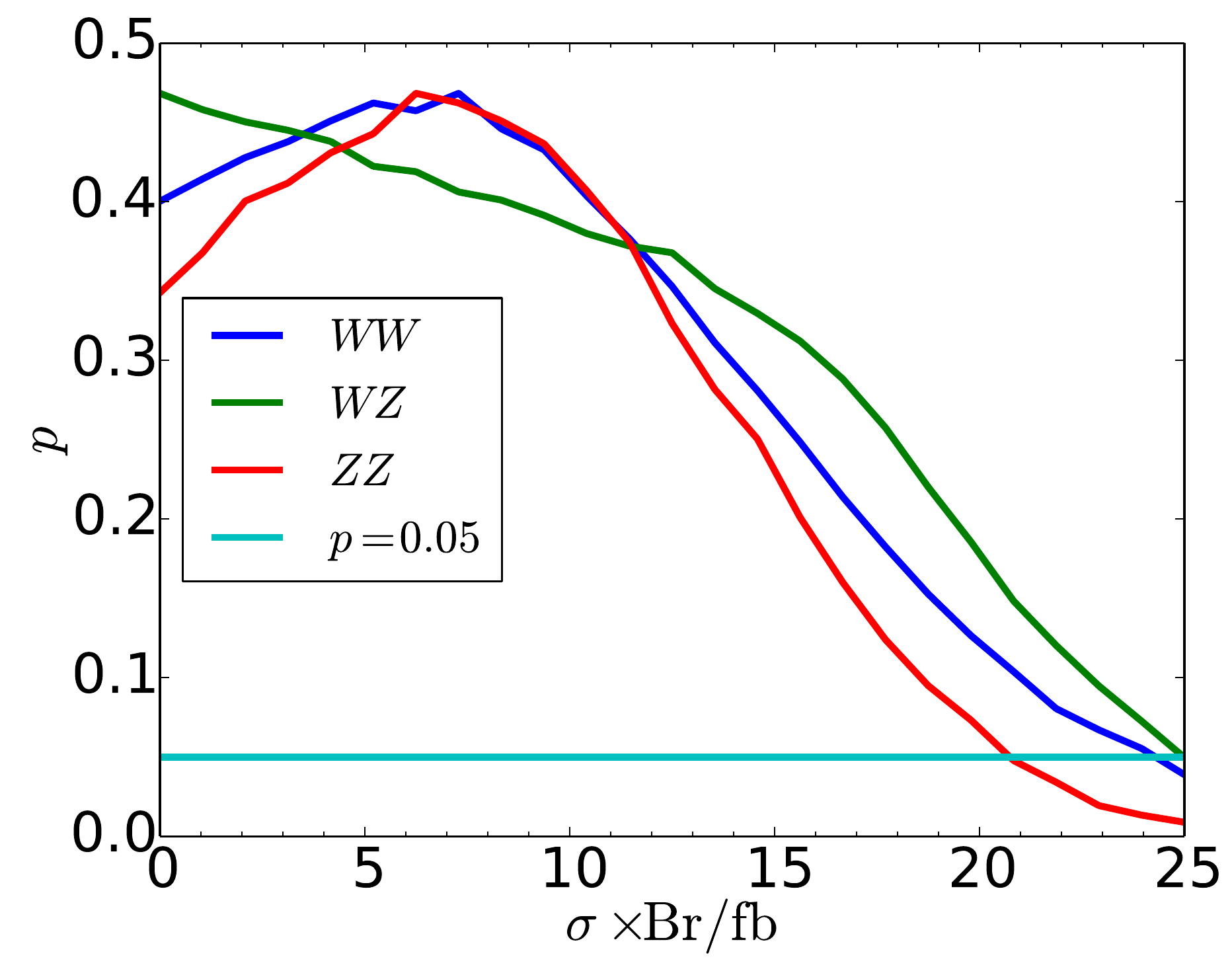}
\caption{\label{fig:oneD} $p-$values as a
  function of production cross section of the 2 TeV resonance $X$ times
  branching ratio for 
each decay channel. The $p-$value has been minimised over the other two signal
regions for each line. 
The horizontal line shows the 95$\%$ 
upper bound. The
efficiencies have been unfolded.}
\end{figure}
\begin{table}
\begin{tabular}{c|ccc}
          & $WW$ & $WZ$ & $ZZ$ \\ \hline
limit/fb  & {\bf 24.3} & {\bf 25.0} & {\bf 20.7} \\
\end{tabular}
\caption{95$\%$ preferred region upper limits on $\sigma(X) \times BR(X
  \rightarrow i)$ 
  coming from the ATLAS fat jets analysis (efficiencies have been
  unfolded).\label{tab:oneD}} 
\end{table}
We now examine the constraints upon each channel individually by
maximising the $p-$value over the other two. 
We show the $p-$values for each individual channel in
Fig.~\ref{fig:oneD}.
In order to find preferred regions of parameter space, we perform $10^4$
pseudoexperiments in order to calculate the $p-$values, maximising the $p-$value
over any unseen dimensions. The 95$\%$ preferred regions (which all have
$p>0.05$) for each channel are shown in 
Table~\ref{tab:oneD}. We see that each channel has an upper bound of around 20
to 25 fb
(equivalent to roughly 400-500 events before efficiencies are taken into
account).  

Within our approximations, the Standard Model for the joint data set
has a $p-$value of 
6$\times 10^{-4}$, equivalent to 4.0$\sigma$ (local significance). This number
of sigma would 
decrease slightly were we to include systematic uncertainties on the
backgrounds, but as stated above: these are rather small and so should not
cause a large effect. We also obtain a larger  local significance than those
quoted by ATLAS because we are combining data rather than analysing
individual channels. 

\begin{figure*}
\includegraphics[width=0.66 \columnwidth]{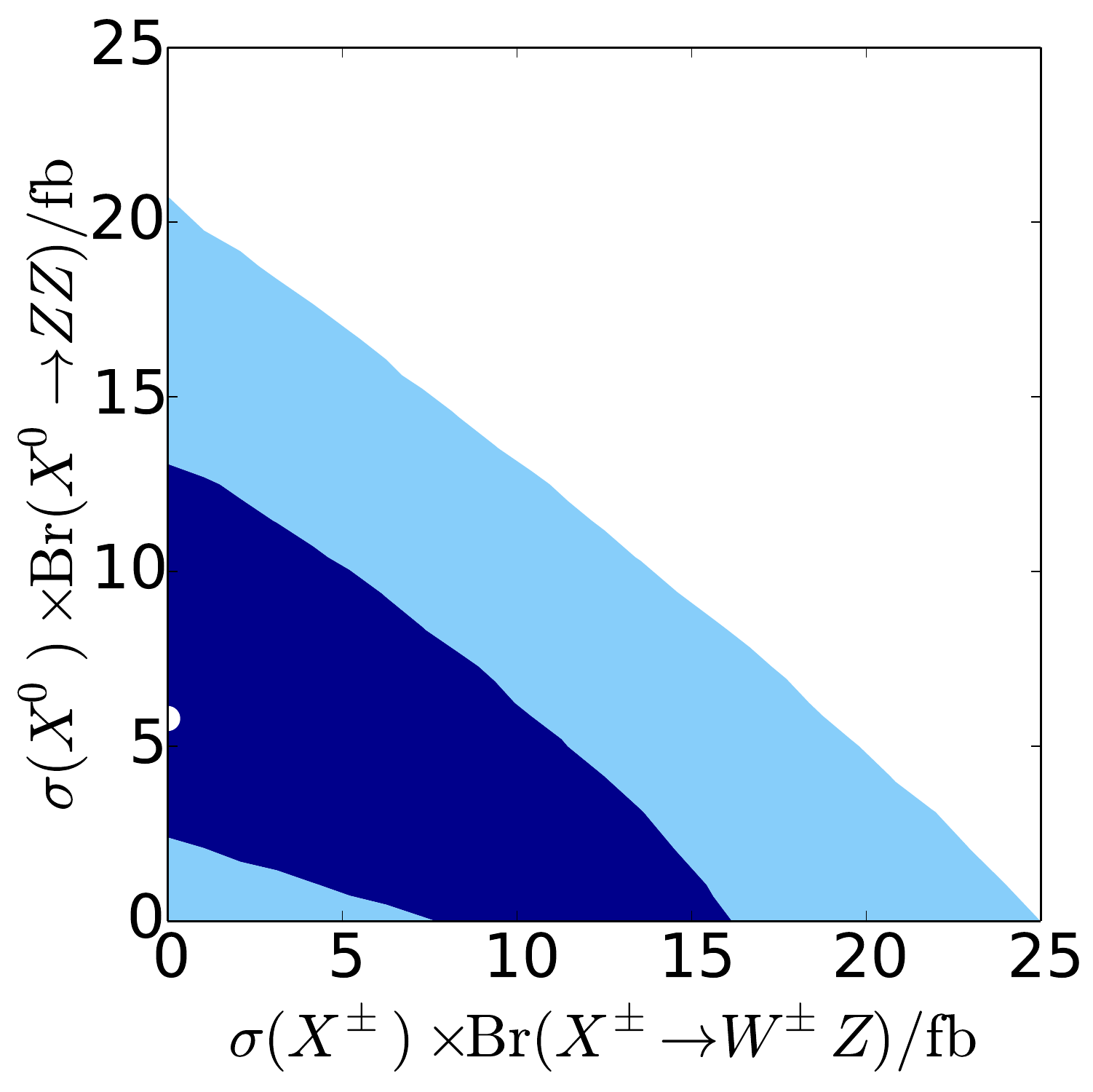}
\includegraphics[width=0.66 \columnwidth]{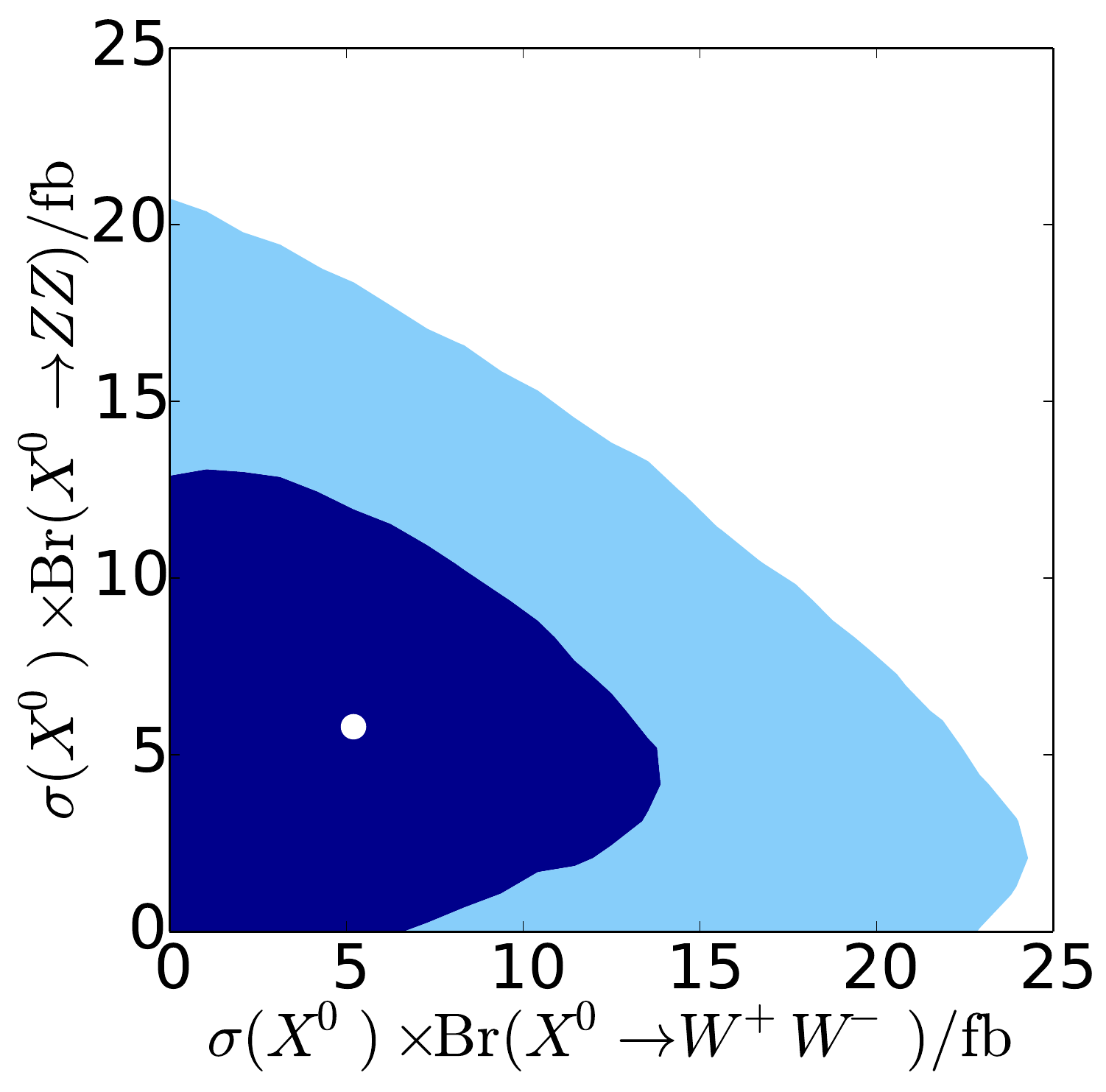}
\includegraphics[width=0.66 \columnwidth]{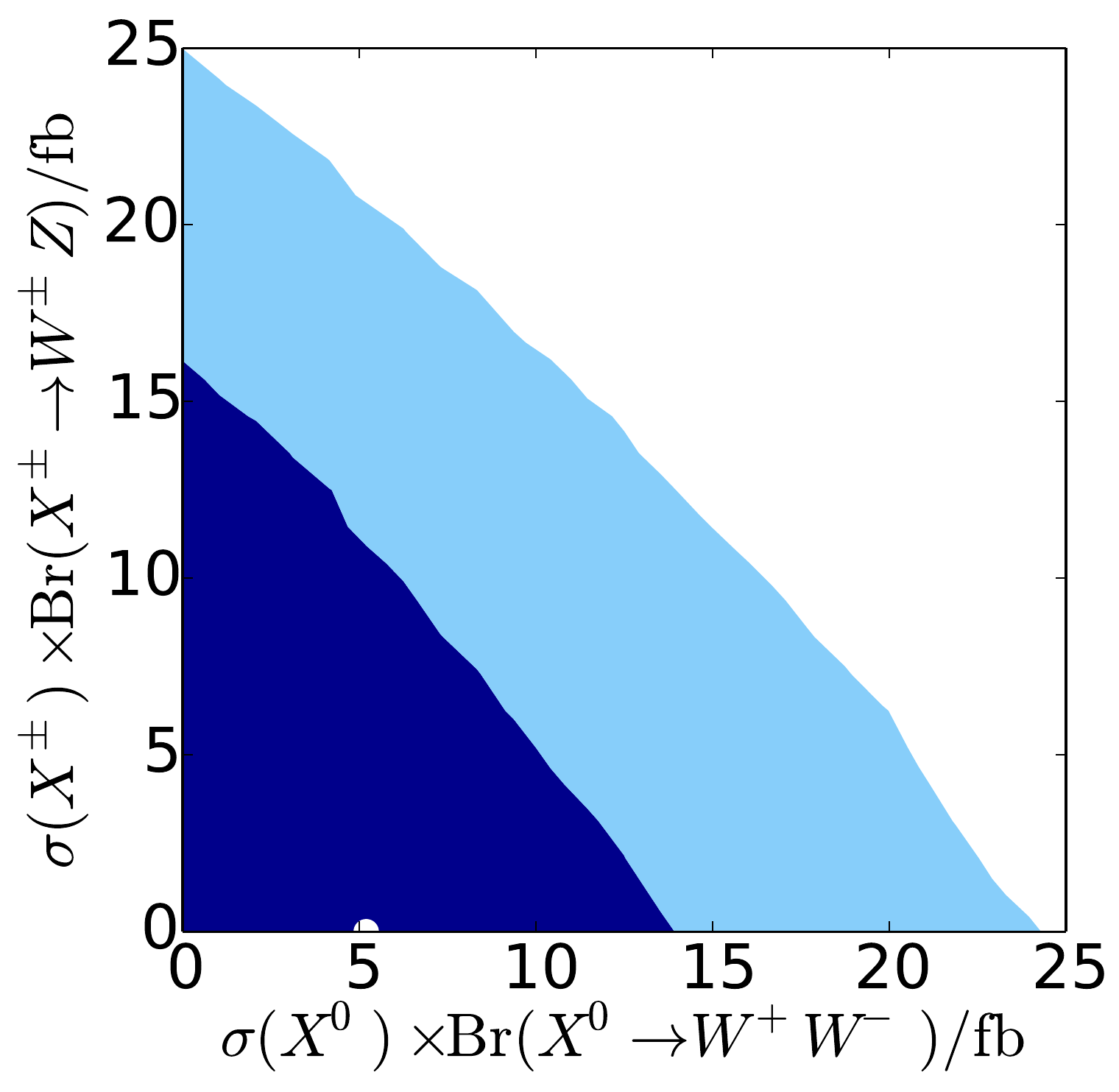}
\caption{Joint constraints on the values of $\sigma \times \mathrm{Br}$
 for different decay channels of a diboson resonance from
 the ATLAS fat jets 
  analysis of the Run I LHC before efficiencies. The 
  darkest 
  region corresponds to 70$\%$ CL, whereas the next darkest region corresponds
to 95$\%$ CL\@. 
In each panel, the best-fit point is denoted by a white dot.\label{fig:joint}}
\end{figure*}
To get joint constraints upon two of the signal channels, we profile over the
unseen one in Fig.~\ref{fig:joint}.
The figure shows that whenever one of the channels has a large $\sigma \times
BR$ (around 20-25 fb), the
anti-correlations imply that the others should be small. The origin is within
the 70$\%$ CL because the unseen $s_j$ is large there, contributing to each of
the 
tagged channels. 
For example, the point
$s_j=\{ 254,\ 0,\ 0\}$ is the best-fit point with $s_{WZ}=s_{ZZ}=0$, i.e.\  
$\sigma(X^0) \times BR(X^0 \rightarrow {W^+W^-})=12.9$ fb,
 $\sigma(X^\pm) \times BR(X^\pm \rightarrow {W^\pm Z})=0$ fb, 
$\sigma(X^0) \times BR(X^0 \rightarrow {ZZ})=0$ fb, predicting expected numbers of diboson tags including SM background
$\mu_{WW}=15.3$, $\mu_{WZ}=15.6$ , $\mu_{ZZ}=5.8$, with $\Delta \chi^2=3.2$
above the best-fit point. 
The unique best-fit point is shown in different projections by
the white dots.

\begin{figure*}
\includegraphics[width=0.66 \columnwidth]{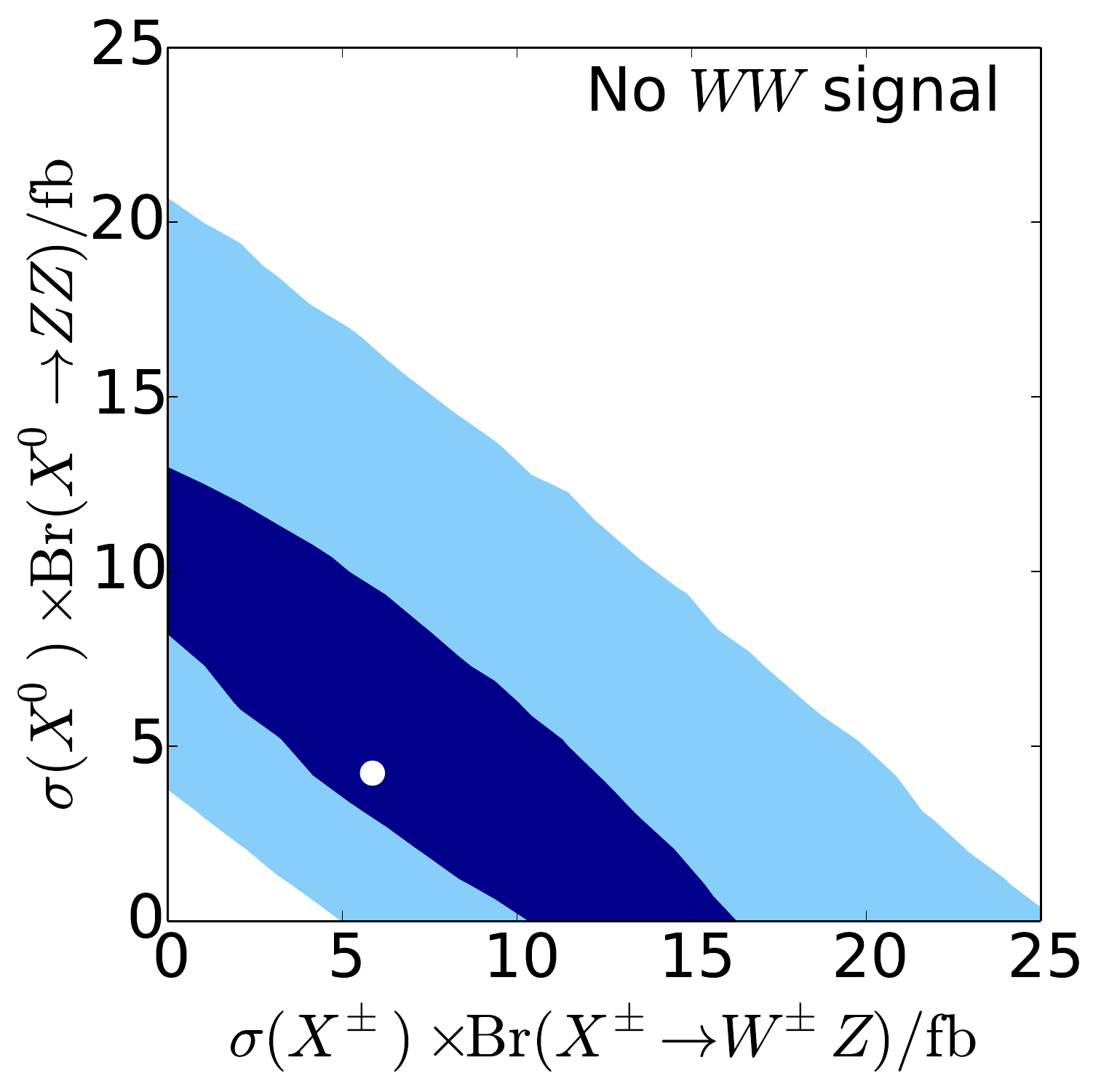}
\includegraphics[width=0.66 \columnwidth]{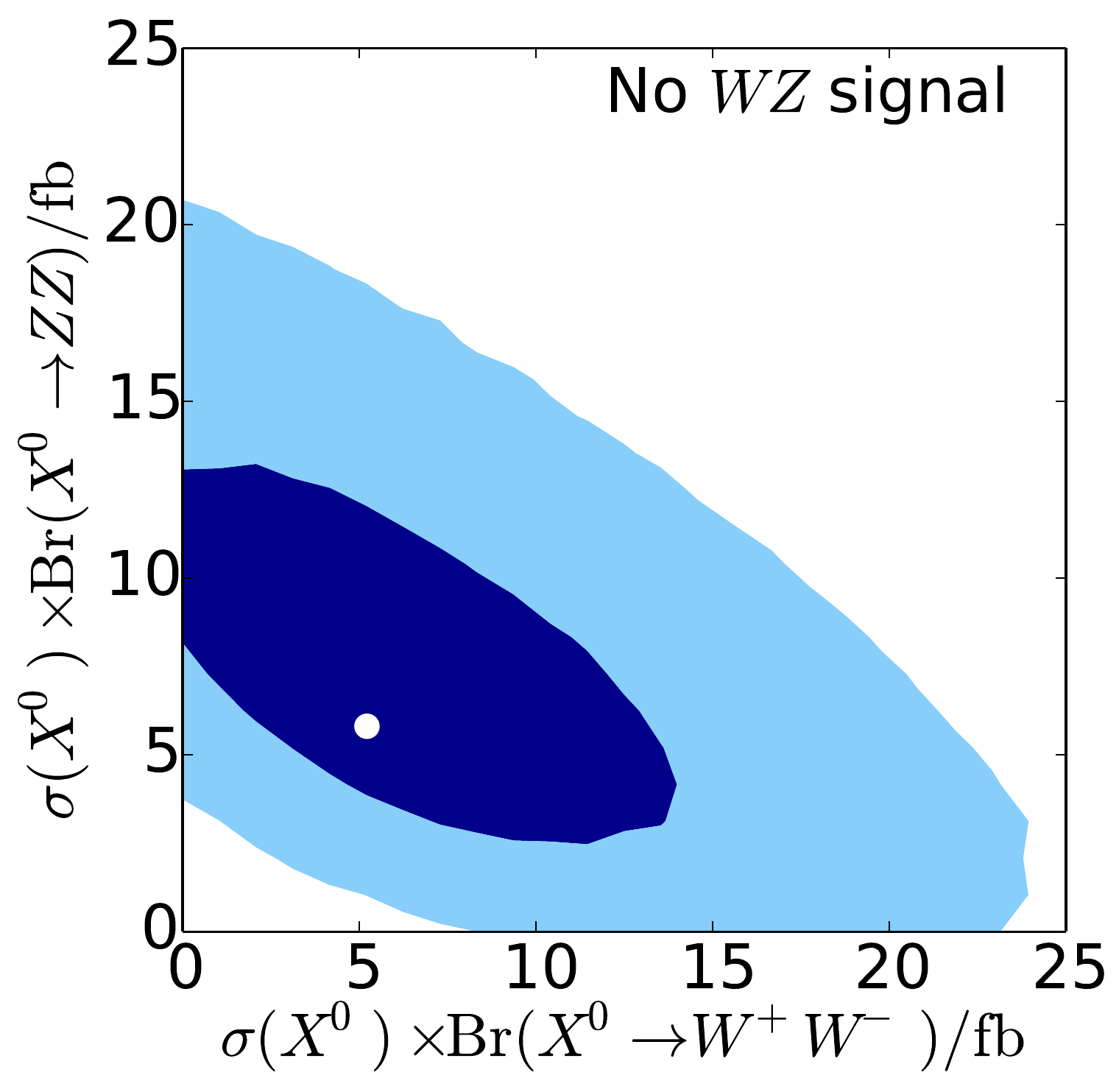}
\includegraphics[width=0.66 \columnwidth]{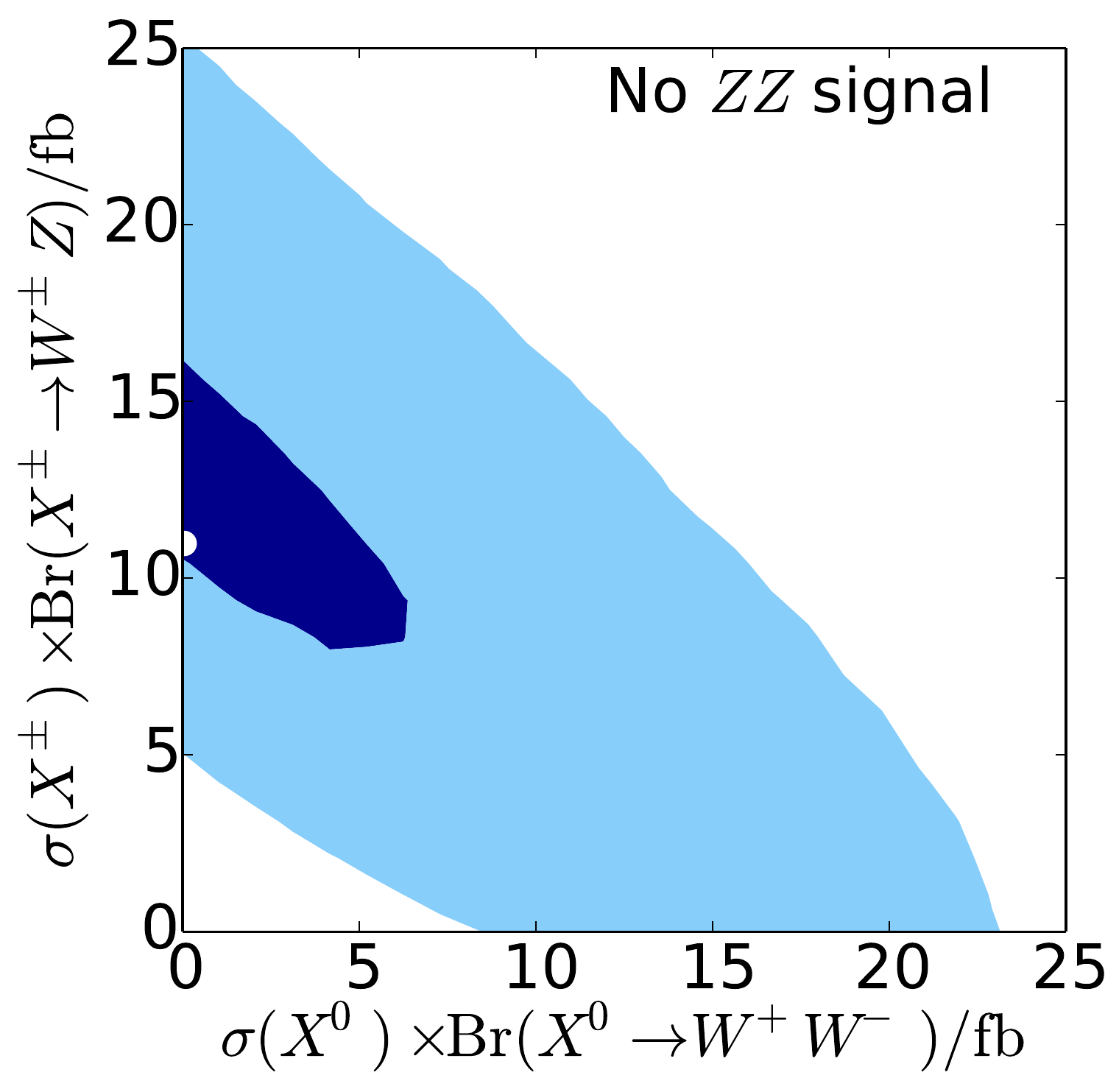}
\caption{Joint constraints on the values of $\sigma \times \mathrm{Br}$
 for different decay channels of a diboson resonance from the ATLAS fat jets
  analysis of the Run I LHC, where one of 
  $s_{WW}$, $s_{WZ}$ or $s_{ZZ}$ is set to
  zero (i.e.\ before efficiency corrections). We show the 70$\%$ and 95$\%$
  preferred regions. In each case, the best-fit point is denoted by a white
  dot.\label{fig:zeroes}}
\end{figure*}
If instead we set one of the $s_j$ to zero (which may be predicted by an
underlying physical model), we obtain the constraints in
Fig.~\ref{fig:zeroes}. 
Now, each panel corresponds to a different model hypothesis, and so unlike
Fig.~\ref{fig:joint}, the best-fit points (displayed by the white points) are
all different.
The three best-fit points are displayed in
Table~\ref{tab:besties}. 
\begin{table}
\begin{tabular}{ccc|ccc|c} 
$s_{WW}$ & $s_{WZ}$ & $s_{ZZ}$ & $\mu_{WW}$ & $\mu_{WZ}$ & $\mu_{ZZ}$ &
$\Delta \chi^2$\\
\hline
{\bf 0} & 119 & 86 & 12.0 & 16.1 & 8.2 & 0.4\\
106 & {\bf 0} & 118 & 13.0 & 16.2 &8.1& 0.0 \\
1 & 223 & {\bf 0} & 13.0 & 16.6 & 7.4& 0.8\\
\end{tabular}
\caption{Best-fit points for the cases where one $s_j$ is set to zero (shown
  in bold).\label{tab:besties}}
\end{table}
We see from the table that each fit has $\Delta \chi^2<1$, meaning that one
cannot significantly discriminate one fit from the other on the basis of ATLAS
fat jets data
alone. 
This situation should improve in future analyses exploiting more sophisticated
jet substructure methods. 

 \begin{figure*}
 \includegraphics[width=0.66 \columnwidth]{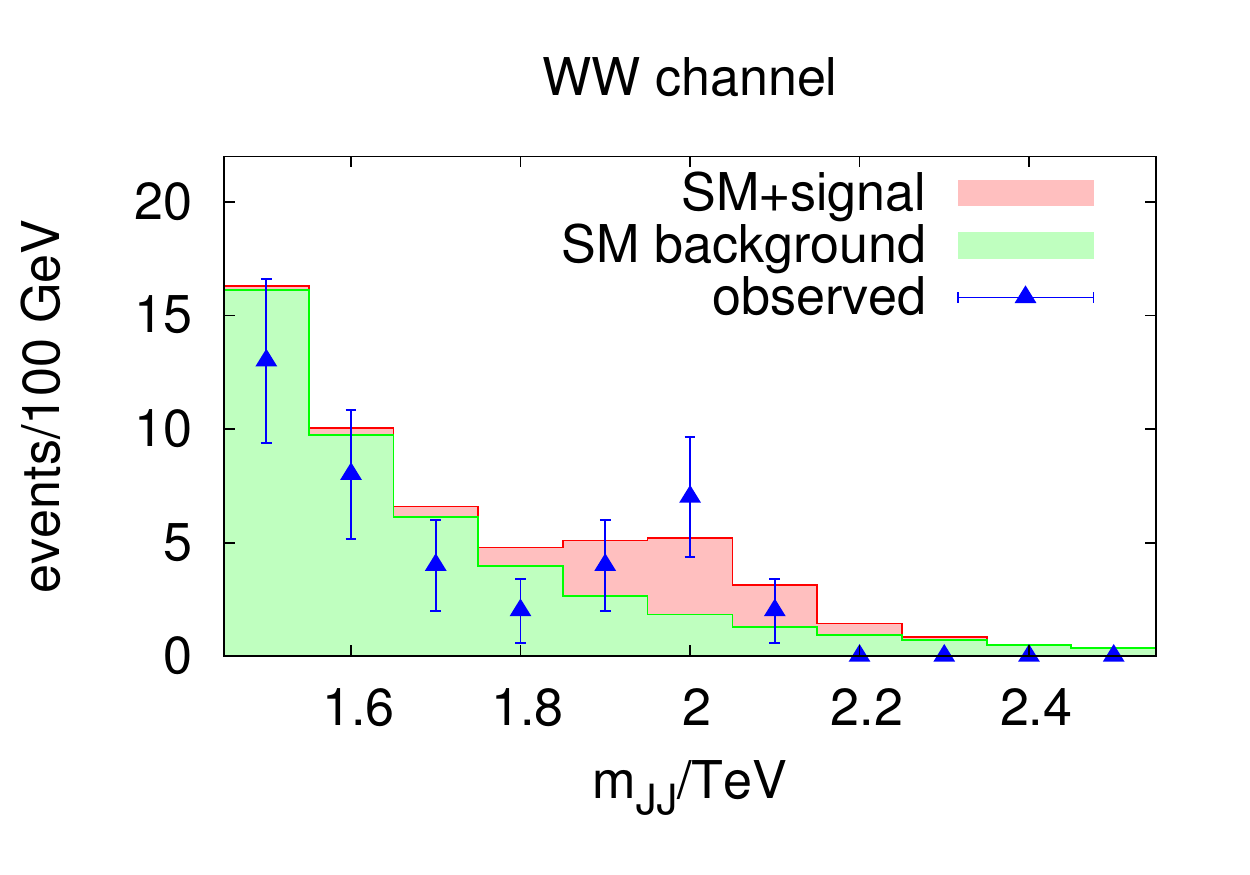}
 \includegraphics[width=0.66 \columnwidth]{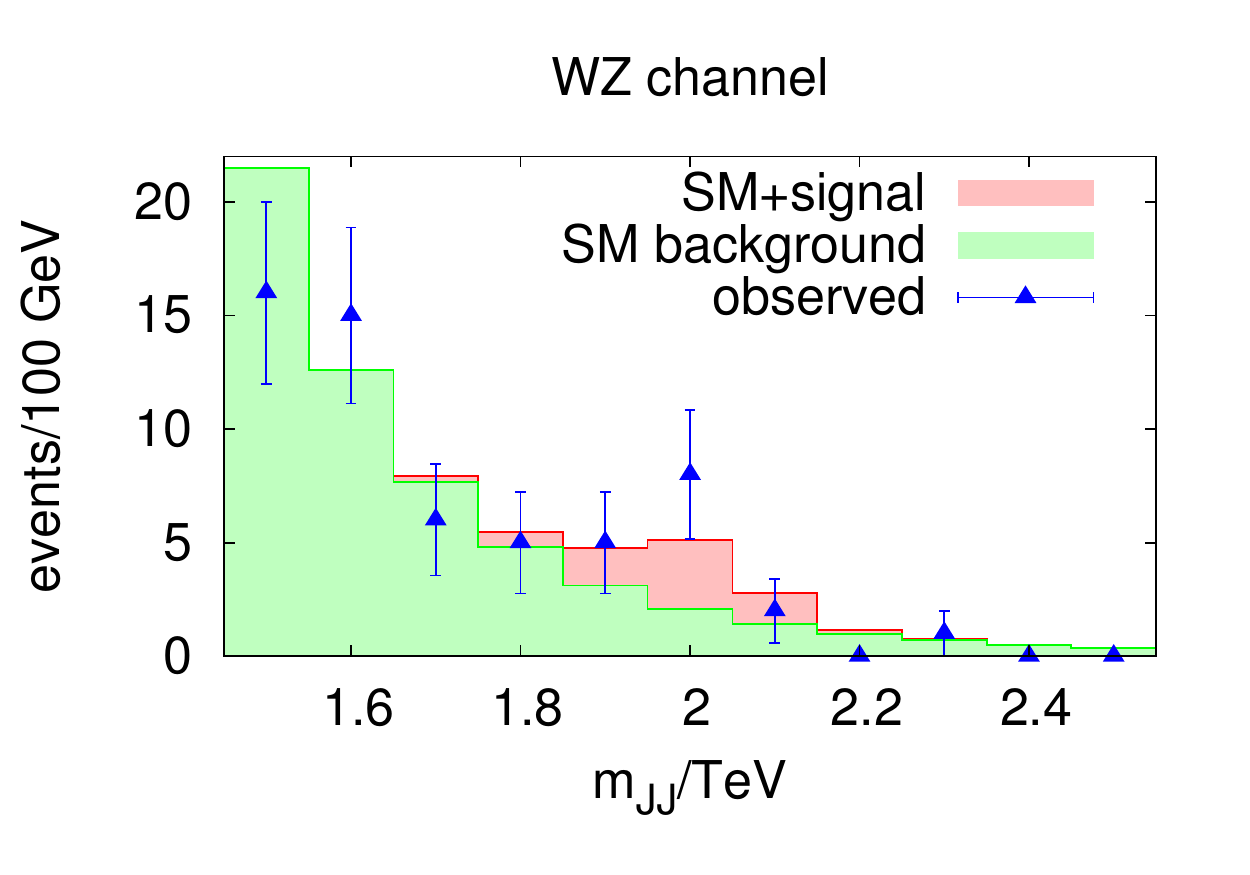}
 \includegraphics[width=0.66 \columnwidth]{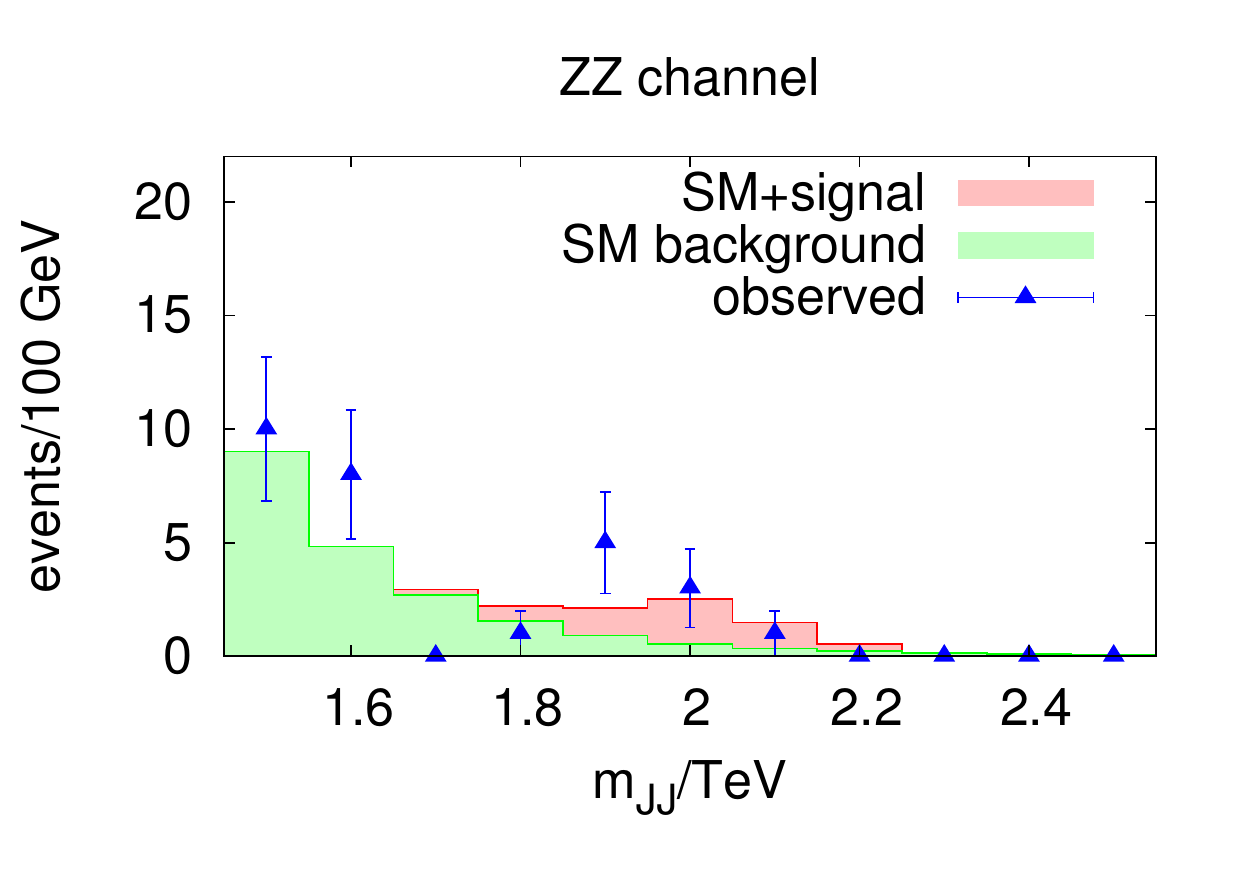}
\caption{Invariant mass distribution near the 2 TeV resonance in each channel
  for $s_{WW}=1$, $s_{WZ}=223$,
$s_{ZZ}=0$. \label{fig:mjj}}
 \end{figure*}
We show the expected jet-jet mass distribution near the 2 TeV
signal region in Fig.~\ref{fig:mjj} for the point $s_{WW}=1$, $s_{WZ}=223$,
$s_{ZZ}=0$ (this corresponds to 
$\sigma(X^0)\times BR(X^0 \rightarrow WW)=0.05$ fb,
$\sigma(X^\pm)\times BR(X^\pm \rightarrow W^\pm Z)=11.0$ fb before
efficiencies), which is the 
best-fit point for $s_{ZZ}=0$: the bottom row of Table~\ref{tab:besties}. The
figure 
shows the contamination in the $WW$ and $ZZ$ channels 
from mis-tagging $WZ$ events. The estimate of the experimental mass resolution
on the resonance was based on those of a 2 TeV $W'$-signal model (whose width is
72 GeV) ATLAS predictions  in 
Ref.~\cite{Aad:2015owa}. The uncertainties placed on the observed numbers of
events are purely statistical ($\sqrt{n}$ for $n$ events), indicating the
expected standard deviation of the measurements. 

Ref.~\cite{Brehmer:2015cia} also performed a likelihood analysis for a
resonance decaying into diboson pairs with similar results. There, a selection
of ATLAS and CMS diboson 
searches are fitted to a wider mass window using a 1.8 TeV resonance rather
than a 2 TeV resonance and so
quantitative differences 
are 
expected, and apparent. We think that it is instructive to examine the
constraints from the 
ATLAS fat jets analysis alone, treating constraints from other diboson
analyses separately.

\section{New physics decalogue \label{sec:model}}
In order to pare down the possible new physics models
explaining the anomaly, we now list a number of qualitative desiderata
for such a model.
\begin{enumerate}[noitemsep,label = (\emph{\roman*})]
\item The discovery of the Higgs boson and measurement of its couplings (as well
as electroweak precision data and flavour physics) all point to
physics being described by 
a theory in which the SM gauge symmetry is spontaneously broken by
the SM Higgs at the weak scale. Unlike some predecessors, we thus
insist that any model respect the SM gauge
symmetry and contain the SM Higgs.

\item The data point to a narrow resonance of high mass (c. 2 TeV). To get a
cross-section times branching ratio in the required range then demands
sizeable couplings in both production (via quarks or gluons) and decay modes. We therefore
insist that these be due to
interactions of dimension four or fewer in the lagrangian.

\item Since the final states are bosonic and there is no evidence for
  the presence of additional invisible particles in the form of
  missing energy, the resonance should have integral
spin $j$. 

\item The requirement of a coupling to gluons or quarks of
dimension $\leq 4$ implies $j \leq 1$.

\item A scalar resonance, $\phi$, with $j=0$
needs electroweak charge 
in order to couple sizeably to light quarks and provide a
production mode. One must ensure both that the scalar does not develop a
vacuum expectation value, which would otherwise, through its Yukawa coupling,
change the masses of the light quarks, and also that the scalar mixes with the
Higgs, facilitating its decay to dibosons. One cannot satisfy both constraints
unless one imposes
{\em ad hoc} relations between different couplings in the Higgs potential. 
Since we are working the context of generic
effective field theories, we wish to avoid such ad hoc relations.

\item A consistent effective field theory (EFT) description of a vector
  resonance 
  $\rho^\mu$, with $j=1$, requires that it be a (massive) gauge field, so we must enlarge
  the SM gauge group somehow.  If $\rho^\mu$ carries electroweak charge, it can couple to both
  quarks and dibosons (possibly via the Higgs field).

\item We require that the couplings preserve the approximate custodial $SU(2)_L \times SU(2)_R$
  symmetry of the SM, both for reasons of economy and because of
  the stringent constraint coming from the electroweak $\rho$ parameter.\footnote{It is possible
    that the couplings required to reproduce the excesses are
    small enough that this requirement can be relaxed. To study this
    requires a detailed electroweak fit for such models, which we leave to future work.}
A coupling to quarks then implies that the resonance transforms as
  either a
  singlet or a triplet of either $SU(2)_L$ or $SU(2)_R$. In the singlet case, however, a coupling to dibosons does
  not result.\footnote{If we allow for custodial symmetry violation,
    the singlet can couple to $WW$.} In the triplet cases, couplings of
the schematic form (we shall be more precise later) $\rho^\mu H^\dagger D_\mu H$ are allowed,
leading to diboson decay modes.

\item A coupling to quarks also yields
  corrections to electroweak precision data that are non-universal, in
  general. At least in the universal limit, with couplings $\lesssim O(1)$, we get tree-level contributions to
  the $S$ parameter (which typically provides one of the strongest constraints) that are acceptably
  small.

\item Sizeable non-universal couplings to quarks also lead to
 corrections to the decay rate of the $Z$ boson to
  hadrons and to the unitarity of the CKM matrix. Such couplings are much less
  constrained if they are to right-handed quarks \cite{Redi:2011zi}, favouring
  the model with a right-handed triplet. One can even exploit
  symmetries to forbid tree-level contributions in this case
  \cite{Agashe:2006at}.
\item In order to avoid problems with flavour physics constraints,
    and for simplicity's sake,
  we assume that the resonance couples in a flavour-diagonal way to
  the two light quark generations only.\footnote{It is likely that this requirement
  can also be
  relaxed somewhat.}
\end{enumerate}
We have thus honed in on a pair of possible models, with either a new $SU(2)_L$ or $SU(2)_R$ triplet
  resonance with sizeable couplings to the Higgs field and light
  quarks. We now build the most general EFTs 
and show that
the anomalies can be explained without contradicting limits on new
physics from other experiments.

\section{EFTs and their fit to data\label{sec:modelfit}}

In this section we write down the most general EFTs satisfying the conditions
of \S\ref{sec:model} (using the rules of \cite{Coleman:1969sm,Callan:1969sn})
and  briefly describe their phenomenology. For each model
we find the parameters that best fit the ATLAS diboson excess.

\subsection{Left handed triplet model\label{sec:lhtripmodelfit}}

Adding a zero-hypercharge heavy vector $SU(2)_L$ triplet $\rho^a_\mu$ 
(indexed by $a \in
\{1,2,3\}$ and comprising three 
charge eigenstates $\rho^+$, $\rho^0$ and $\rho^-$) to the SM results in the most general lagrangian up to dimension
four of
\begin{align*}
\lag = \lag_\text{SM} -\frac{1}{4} \rho^a_{\mu \nu} \rho^{a \mu \nu} + (\frac{1}{2} m_\rho^2 +\frac{1}{4} g_m^2  H^\dagger H) \rho^a_\mu \rho^{a \mu} \\
- 2 g \epsilon^{abc} \pt_{[\mu} \rho^a_{\nu]} W^{b \mu} \rho^{c \nu} 
- g \epsilon^{abc} \pt_{[\mu} W^a_{\nu]} \rho^{b \mu} \rho^{c \nu} \\
+ (\frac{1}{2} i g_\rho \rho^a_\mu H^\dagger \sigma^a D^\mu H + \text{h.c.}) + g_q \rho^a_\mu \overline{Q_L} \gamma^\mu \sigma^a Q_L \\
+ g_l \rho^a_\mu \overline{L_L} \sigma^a \gamma^\mu L_L + \ldots, 
\end{align*}
where $\sigma^a$ are the Pauli matrices, $g$ is the $SU(2)_L$ gauge
coupling and $\partial_{[\mu } \rho^a_{\nu  ]} \equiv
\frac{1}{2}(\partial_\mu \rho^a_\nu - \partial_\nu \rho^a_\mu)$. \footnote{A
  similar lagrangian was considered in 
  Ref.~\cite{Zerwekh:2005wh}.} 
The coefficient of the term $-\epsilon^{abc} \pt_{[\mu} W^a_{\nu]} \rho^{b \mu}
\rho^{c \nu}$ is set to $g$ because such a value results in a higher ultra-violet
cut-off scale $\Lambda$, where $\Lambda$ is associated with unitarity
violation. However, one could also consider small deviations from $g$ of order
$gm_\rho^2/\Lambda^2$: these must not be large otherwise the r\'{e}gime of
validity of our EFT is compromised.
There are additional terms that we have not written, such as $\rho^2 W^2$,
that do 
  not affect the discussion here, but which restore $SU(2)_L$ gauge invariance
  and may be relevant for future
  searches.
The `$\rho H^\dagger D H$' coupling, after electroweak symmetry breaking
(EWSB), mixes the $\rho^\pm$ 
with the $W^\pm$, and the $\rho^0$ with the $Z$, with mixing angle of
order $\frac{g g_\rho v^2}{4 m_\rho^2} $ for Higgs vacuum expectation value $v = 246 \Gev$,
analogously to the rho meson in hadronic physics. The same operator mediates the decay of the $\rho^0$ to $W^+ W^-$ and $Z h$, and that of the $\rho^\pm$ to $W^\pm Z$ and $W^\pm h$.

\begin{figure}
\includegraphics[width=\columnwidth]{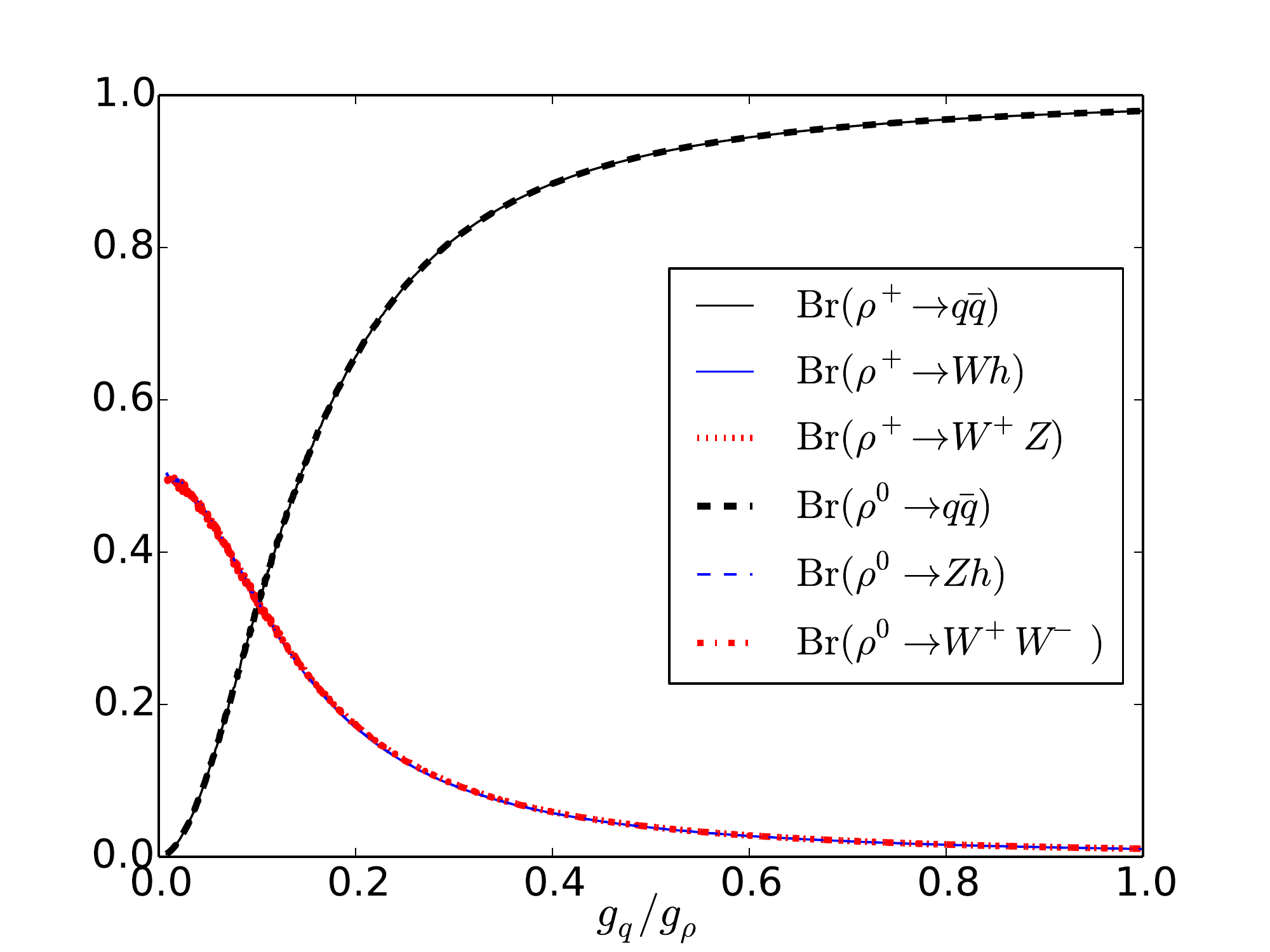}
\caption{\label{fig:leftbrs} The branching ratios of the $\rho^+$ and $\rho^0$
  of the left handed triplet model, as a function of their coupling to quarks
  $g_q$ over their coupling to bosons $g_\rho$. Note the equal branching
  ratios to $Wh$ and $WZ$, and also to $Zh$ and $WW$, as predicted by
  \cite{Thamm:2015csa}. The dijet branching ratios of the $\rho^+$ and
  $\rho^0$ overlap (black curves), as do their diboson branching ratios (blue
  and red curves).} 
\end{figure}

\begin{figure}
\unitlength=\columnwidth
\begin{picture}(1,0.75)
\put(-0.2,.8){\includegraphics[angle=270,width=1.2 \columnwidth]{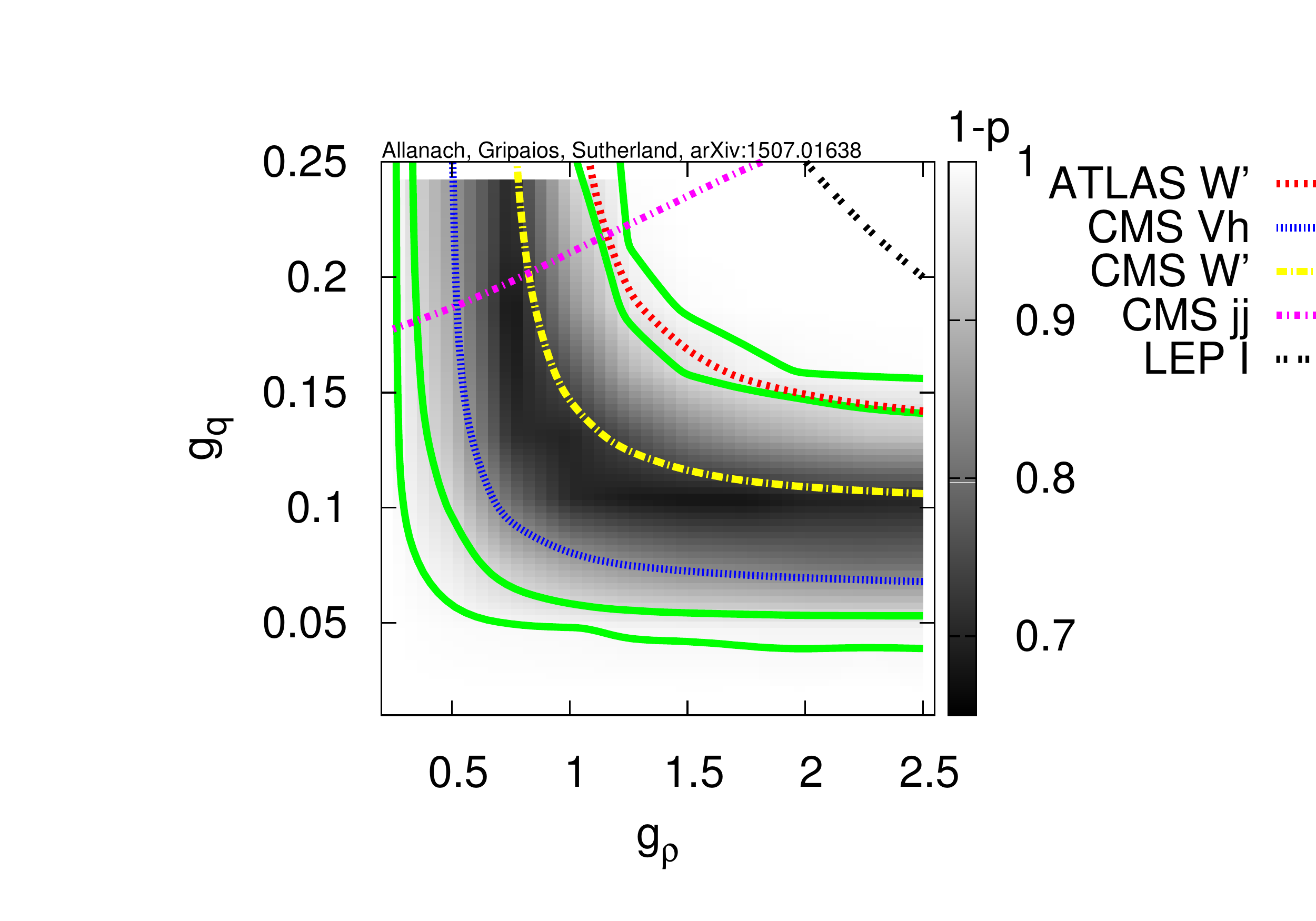}}
\end{picture}
\caption{\label{fig:lefttripletchisq} Preferred
  regions in the plane of the $SU(2)_L$ triplet's coupling to bosons,
  $g_\rho$, and quarks, $g_q$, as determined from the number of events
  observed in the six disjoint signal regions (see
  Table~\ref{tab:eventNumbers}). We show the 95$\%$, 99$\%$ preferred regions
  by the 
  inner and outer pair of solid lines, respectively.
  Also shown are the 95\% CL limits on a
  $W^\prime$ 
  model from \cite{Aad:2014xka} (ATLAS W') and \cite{Khachatryan:2014hpa}
  (CMS W') which should
  be similar to the limits on the $SU(2)_L$ triplet considered here, the
  limit from the CMS search for resonances decaying to $Wh,Zh\to qqb\bar{b}$ 
  \cite{Khachatryan:2015bma} (`CMS Vh'), and the limit from the CMS dijet
  resonance 
  search~\cite{Khachatryan:2015sja} (`CMS jj'). The line denoted `LEP I'
  depicts the 
  approximate   constraint 
  from electroweak precision tests at LEPI\@. The region above each broken line is
  excluded.}
\end{figure}

As described above, we assume
the $\rho$ only couples to the first two quark generations, with equal strength; we also set $g_l = 0$, given the absence of a $2 \Tev$ bump in dilepton searches \cite{Khachatryan:2014fba,Aad:2014cka}. We assume for simplicity that $g_m=0$; for example a $g_m=4$ would only increase the partial width of the $\rho$s to either $Wh$ or $Zh$ by $\sim 10\%$. We use FeynRules 2.0.6~\cite{Alloul:2013bka} and MadGraph5\_aMC@NLO
v2.2.3~\cite{Alwall:2014hca} to simulate the production and decay of
the $\rho$s at leading order, using a $K$ factor of 1.3 consistent with that
of Drell-Yan $W^\pm$ production \cite{Anastasiou:2003ds}. The number of
produced $\rho$s are then multiplied by their branching ratios to dibosons
(Fig.~\ref{fig:leftbrs}), the 
efficiencies in Table~\ref{tab:dib} and the overall efficiency factor of
$\epsilon$ to
obtain a prediction for the number of signal events in the six disjoint regions $A$ to $F$, as a function of the lagrangian parameters. Using the 
observations of Table~\ref{tab:eventNumbers}, we perform
pseudoexperiments to obtain a $p$ value for
each set of parameters; Fig.~\ref{fig:lefttripletchisq} shows the
resulting good-fit regions in the
$(g_\rho,g_q)$ plane.
 Towards the top of the best fit region the
$\rho$s are produced copiously but rarely decay to dibosons, whereas
towards the right the $\rho$s are produced rarely but almost always
decay to dibosons (where `dibosons' includes the decays to $Wh$ or $Zh$). We also overlay in Fig.~\ref{fig:lefttripletchisq} the 95\%
CL limits on $\sigma \times \text{Br}(W^\prime \to WZ)$ from other
searches for diboson resonances, namely the CMS all hadronic search
\cite{Khachatryan:2014hpa} ($12 \, \mathrm{fb}$) and the ATLAS
semileptonic search \cite{Aad:2014xka} ($20 \, \mathrm{fb}$). Note
that we do not consider the CMS semileptonic search, because the only
readily available limits are for a type I RS graviton, which has
considerably higher acceptances than, say, a $W^\prime$. Given the
similarity of the ATLAS and CMS semileptonic limits on the type I RS
graviton, we assume any recasting of the CMS search onto the triplet model of
this section would yield limits comparable to the ATLAS $W^\prime$ limit
displayed in Fig.~\ref{fig:lefttripletchisq}. 

Also shown in Fig.~\ref{fig:lefttripletchisq} is the CMS 95\% CL limit
on $\sigma(X) \times \text{Br}(X \to W h, Z h) = 8 \, \mathrm{fb}$ for
a $2 
\Tev$ spin one resonance $X$ \cite{Khachatryan:2015bma}. The limit
is quite constraining for the $SU(2)_L$ triplet, given the roughly equal
branching ratio of the 
$\rho^\pm$ to $WZ$ and $Wh$, as well as that of the 
$\rho^0$ to $WW$ and $Zh$. 
Interestingly, the analogous limit for a marginally lighter $1.8
\Tev$ resonance is much weaker ($14 \, \mathrm{fb}$).

We now comment on the compatibility with electroweak precision
  constraints.\footnote{Electroweak fits to similar were performed in \cite{delAguila:2010mx,deBlas:2012qp}, but
    do not lead to significant constraints on the models considered here.} The model is non-universal, but we can estimate the constraints by assuming that
    $\rho$ couples equally to all 3 quark generations, such that we may compare with the analysis performed
   using a flavour-symmetric basis of dimension-six SM operators in \cite{Pomarol:2013zra}. Integrating
  out the $\rho$, we obtain 3 such operators:
  $\frac{g_\rho^2}{4 m_\rho^2} (i H^\dagger \sigma^a  \lra{D_\mu} H)
  (i H^\dagger \sigma^a  \lra{D^\mu} H)$, $\frac{g_\rho g_q}{2
    m_\rho^2} \mathcal{O}^{(3)q}_L \equiv \frac{g_\rho g_q}{2
    m_\rho^2} (i H^\dagger \sigma^a  \lra{D_\mu} H) (\overline{Q_L}
  \sigma^a \gamma^\mu Q_L)$, and $\frac{g_q^2}{m_\rho^2}
  (\overline{Q_L} \sigma^a \gamma_\mu Q_L) (\overline{Q_L} \sigma^a
  \gamma^\mu Q_L)$. Re-writing these in the basis of
    \cite{Elias-Miro:2013mua}, we find that only
  $\mathcal{O}^{(3)q}_L$, contributes to Z pole measurements. We use the 95\% CL limit on its Wilson
  coefficient alone, given in Eq.~(19) of \cite{Pomarol:2013zra}, to place the
  approximate bound $\abs{g_\rho g_q} \lesssim 0.5$, which is
  compatible with the values required to fit the excess in the ATLAS
  diboson search (see the grey dashed line in Fig.~\ref{fig:lefttripletchisq}).

The $\rho$ boson necessarily
  couples to 
quarks (in order to obtain the production cross-section), and so we should
consider constraints coming from resonance searches to dijets at an invariant
mass of 2 TeV. CMS, for instance, places a 95$\%$CL upper limit of 60
fb~\cite{Khachatryan:2015sja} for 
$\sigma \times BR(\rho \rightarrow q \bar q)\times A$, where $A\leq 1$ is
acceptance (ATLAS' analogous upper bound is 110 fb~\cite{Aad:2014aqa}). 
Assuming an acceptance $A \sim 0.6$, as quoted in \cite{Khachatryan:2015sja} for isotropic decays, the CMS limit rules out the otherwise good
fit points with large $BR(\rho \rightarrow q \bar q)$, as shown in
Fig.~\ref{fig:lefttripletchisq}, preferring instead a sizeable branching ratio of the $\rho$ to dibosons.

\subsection{Right handed triplet model}

Applying the same logic as in section~\ref{sec:lhtripmodelfit}, the most
general 
lagrangian up to dimension four containing an additional triplet of
$SU(2)_R$, $\rho_\mu^a$, is\footnote{We have neglected a small
    mass splitting, of $O(\frac{g^{\prime 2}}{g_\rho^2})$, in $m_\rho$.}
\begin{align*}
\lag = \lag_\text{SM} -\frac{1}{4} \rho^a_{\mu \nu} \rho^{a \mu \nu} + (\frac{1}{2} m_\rho^2 +\frac{1}{4} g_m^2  H^\dagger H) \rho^a_\mu \rho^{a \mu} \\
- 2 g^\prime \epsilon^{ab3} \pt_{[\mu} \rho^a_{\nu]} \rho^{b \mu} B^{\nu}  - g^\prime \epsilon^{3bc} \pt_{[\mu} B_{\nu]} \rho^{b \mu} \rho^{c \nu} \\
+ (-\frac{1}{4} i g_\rho \rho^a_\mu \Tr( \Pi \sigma^a D^\mu \Pi^\dagger) + \text{h.c.}) + g_q \rho^a_\mu \overline{Q_R} \gamma^\mu \sigma^a Q_R ,
\end{align*}
where $Q_R = \begin{pmatrix} u_R \\ d_R \end{pmatrix}$, $g^\prime$ is the
$U(1)_Y$ gauge coupling, and we have taken advantage of notation in which the
$SU(2)_L\times SU(2)_R$ symmetry of the Higgs doublet $H = \begin{pmatrix}
  \phi^+ \\ \phi^0 \end{pmatrix}$ is manifest, defining $\Pi = (H,H^c)
= \begin{pmatrix} \phi^+ & \overline{\phi^0} \\ \phi^0 & -
  \phi^- \end{pmatrix}$ and $D_\mu \Pi = \pt_\mu \Pi + \frac{1}{2} i g W^a_\mu
\sigma^a \Pi + \frac{1}{2} i g^\prime B_\mu \Pi \sigma^3$. Much the same
phenomenology results as in the left handed triplet case: the charged and
neutral components of the $\rho$ mix to the same degrees with the $W$s and $Z$
respectively (after EWSB); they can also decay to $WZ$/$Wh$, or $WW$/$Zh$,
respectively.

\begin{figure}
\unitlength=\columnwidth
\begin{picture}(1,0.75)
\put(-0.2,.8){\includegraphics[angle=270,width=1.2 \columnwidth]{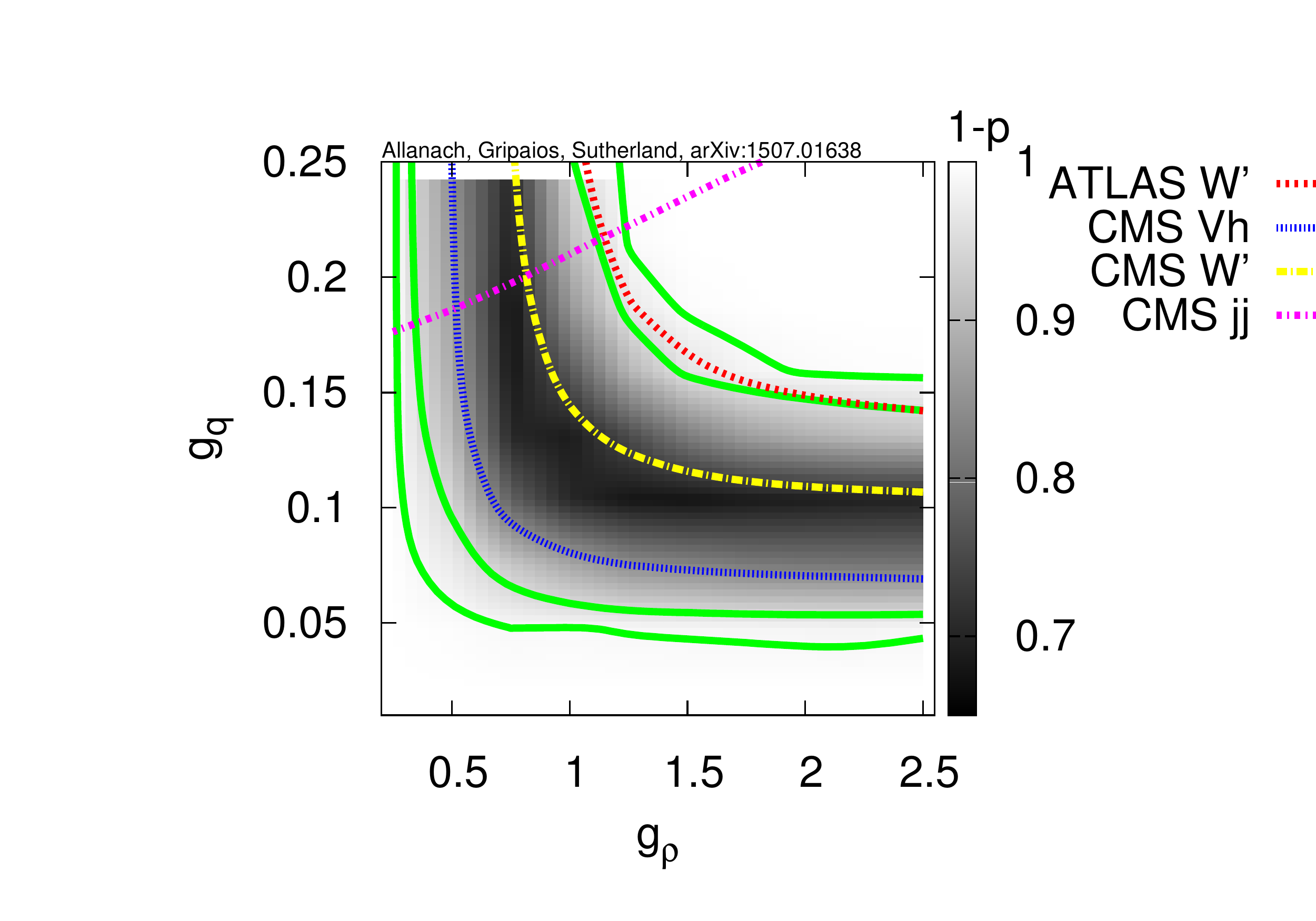}}
\end{picture}
\caption{\label{fig:righttripletchisq} Preferred
  regions in the plane of the $SU(2)_R$ triplet's coupling to bosons,
  $g_\rho$, and quarks, $g_q$.  We show the 95$\%$, 99$\%$ preferred regions
  by the 
  inner and outer pair of solid lines, respectively. 
  Also shown are the 95\% CL limits on a
  $W^\prime$ 
  model from \cite{Khachatryan:2014hpa} (ATLAS W') and \cite{Aad:2014xka}
  (CMS W') which should
  be similar to the limits on the $SU(2)_R$ triplet considered here, the
  limit from the CMS search for resonances decaying to $Wh,Zh\to qqb\bar{b}$ 
  \cite{Khachatryan:2015bma} (`CMS Vh'), and the limit from the CMS dijet
  resonance 
  search~\cite{Khachatryan:2015sja} (`CMS jj').  The region above each broken
  line is   excluded.}
\end{figure}

The branching ratios of the $\rho$s are identical to those of the left-handed
triplet model, shown in Fig.~\ref{fig:leftbrs}. An
identical analysis to \S\ref{sec:lhtripmodelfit} yields 
Fig.~\ref{fig:righttripletchisq}, showing the points in the $(g_\rho,g_q)$
plane that best fit the ATLAS diboson excess, along with relevant constraints
from other diboson resonance searches. A comparison with
Fig.~\ref{fig:lefttripletchisq} shows that the fit to the diboson
anomaly is practically identical to the $SU(2)_L$ triplet. The other
constraints are also identical, except for the EWPT. 

Unfortunately, we cannot perform a robust fit to EWPT using
  \cite{Pomarol:2013zra} in this case, because integrating out the
  $\rho$ generates operators such as $(i \tilde
  H^\dagger \lra{D_\mu} H) (\overline{u_R} \gamma^\mu d_R)$ that are
  not considered there. So a detailed fit {\em ab initio}\/ is required,
  which we leave for future work. As we argued above, the constraints
  will be much
weaker in this case, because contributions to hadronic decays of the
$Z$ are suppressed (typically by an order of magnitude
\cite{Redi:2011zi,Pomarol:2013zra}) and because CKM unitarity
violation is absent. We thus expect that there will be no significant
constraint on the region of allowed couplings.

\section{Conclusions}

Figs.~\ref{fig:lefttripletchisq} and~\ref{fig:righttripletchisq} show that
right- or left- handed triplets can explain 
the ATLAS 
diboson excesses without contradicting other constraints. 
Our effective field theory analysis should be less model dependent than
specific models that have recently appeared in the literature.
Indeed, we provide general likelihood constraints on a resonance which can
decay via channels $WW$, $WZ$ and $ZZ$ in
Figs.~\ref{fig:joint},\ref{fig:zeroes}. 
The production cross sections of both the $SU(2)_L$ and $SU(2)_R$ triplets
increase by a factor of $7$ in $13 \Tev$ collisions (relevant for Run II of
the LHC), compared to those at $8
\Tev$ on which the ATLAS fat-jets diboson resonance search was based. By the
end of 2015, the models considered would conservatively 
predict at least as many signal events as in previous run, which would be
observed foremost in the $WZ$ and $Wh$ all-hadronic channels. The
  channels where the $W$s and $Z$s decay leptonically are presently a factor
  of $\sim 2$ less sensitive; this may improve if the efficiency of jet
  substructure methods worsens due to higher pileup.

\section*{Acknowledgements}
This work has been partially supported by STFC grant 
ST/L000385/1. 
We thank M.~Redi, J.~Tattersall, J.~Thaler, A.~Wulzer and members of the
Cambridge SUSY Working Group for 
discussions. BG acknowledges
MIAPP and King's College,
Cambridge. DS
acknowledges the support of Emmanuel College, Cambridge.
\bibliographystyle{apsrev}
\bibliography{diboson}

\end{document}